\documentclass[twocolumn,showpacs,aps]{revtex4}
\usepackage{amssymb}
\newcommand{\half}{\frac{1}{2}}
\newcommand{\thrd}{\frac{1}{3}}
\newcommand{\frth}{\frac{1}{4}}
\newcommand{\sxth}{\frac{1}{6}}
\begin{document}
\title{
Higgs mass determined by cosmological parameters }
\author{R. K. Nesbet }
\affiliation{
IBM Almaden Research Center,
650 Harry Road,
San Jose, CA 95120-6099, USA }
\date{\today}
\begin{center}For {\em PhysRevLett} \end{center}
\begin{abstract}
Postulating that all massless elementary fields have conformal scaling
symmetry removes a conflict between gravitational theory and the 
standard model of elementary quantum fields.  If the scalar field 
essential to SU(2) symmetry breaking has conformal symmetry, it must 
depend explicitly on the Ricci curvature scalar of gravitational theory.
This has profound consequences for both cosmology and elementary 
particle physics, since cosmological data determine scalar field 
parameters.  A modified Friedmann equation is derived and solved
numerically.   The theory is consistent with all relevant data for
supernovae redshifts below $z=1$.  The implied value of the 
cosmological constant implies extremely small Higgs mass, far below 
current empirical lower bounds.  Detection of a Higgs boson with large 
mass would falsify this argument.
\end{abstract}
\pacs{04.20.Cv,11.15.-q,14.80.Bn}
\maketitle
\section{Introduction}
Einstein gravitational theory has lower symmetry than the standard model
of spinor and gauge boson fields.  Massless electroweak action integrals
are invariant under local Weyl (conformal) scaling, 
$g_{\mu\nu}(x) \to g_{\mu\nu}(x) e^{2\alpha(x)}$,
for fields with definite conformal character.  For example,
scalar field $\Phi(x)\to\Phi(x)e^{-\alpha(x)}$.   A conformal  
energy-momentum 4-tensor is traceless, while the Einstein tensor is not.
Compatibility can be imposed by replacing the
Einstein-Hilbert field action by a uniquely determined action integral
constructed using the conformal Weyl tensor, which preserves general 
relativistic phenomenology at the distance scale of the solar 
system\cite{MAN06}.
Discrepancies at galactic distances are commonly attributed 
to unobserved dark matter.  Most remarkably, conformal theory
is consistent with the empirically successful MOND
model\cite{MIL83} in describing the systematics of these discrepancies.
This provides an alternative explanation of excessive rotational
velocities in the outer regions of galaxies, without invoking dark
matter\cite{MAN06}.
\par In the electroweak standard model\cite{PAS95}, mass is generated
by an SU(2) doublet complex scalar field $\Phi$.  
The Lagrangian density contains 
$\Delta{\cal L}_\Phi=w^2\Phi^\dag\Phi-\lambda(\Phi^\dag\Phi)^2$,
where $w^2$ and $\lambda$ are positive constants.
The Higgs mass is $m_H=\sqrt{2}w$.
Units here are such that $\hbar=c=1$.
$\lambda(\Phi^\dag\Phi)^2$ is conformally covariant, but conformal 
symmetry is broken by $w^2\Phi^\dag\Phi$.  
The Higgs construction breaks SU(2) and conformal symmetries by setting 
$\Phi^\dag\Phi=\phi^2_0$, for spacetime constant $\phi_0$.  Conformal
theory replaces the $w^2$ term by $-\sxth R\Phi^\dag\Phi$, where 
scalar $R=g_{\mu\nu}R^{\mu\nu}$ for Ricci tensor $R^{\mu\nu}$, the
symmetric contraction of the gravitational Riemann tensor\cite{MAN06}.
However, empirical cosmological parameters indicate that $R>0$.
This sign conflict is resolved here by including both terms, while  
$\phi_0$ is modified to include the effect of $w^2-\sxth R$.
The residual constant term in the Lagrangian density defines 
an effective cosmological constant ${\bar\Lambda}$\cite{MAN06}.   
\par If $w^2$ is a dynamical result of self-interaction
\cite{NES08}, it may be much smaller than 
commonly assumed.  Conformal covariance of the bare electroweak and
gravitational theories establishes a relationship between parameters
$w^2$ and ${\bar\Lambda}$.  The latter
is currently known from empirical cosmology, implying a very small
value of $w^2$ and Higgs mass $m_H$ many
orders of magnitude smaller than empirical limits inferred using 
the standard model\cite{GHKD90}.  However, important mechanisms of
Higgs production and detection would be removed if Higgs-fermion
coupling were eliminated\cite{NES08}, which would require 
reconsideration of theoretical expectations.  In any case, detecting 
a spinless neutral particle with mass less than the electron
neutrino presents a great experimental challenge.

\section{Conformal theory including a scalar field}
For homogeneous isotropic (Robertson-Walker) geometry, 
the conformal gravitational term vanishes in the coupled field equations,
requiring the total source energy-momentum tensor to vanish\cite{MAN06}.
However, the Ricci scalar in the Lagrangian density of a conformal 
scalar field produces a gravitational term in its energy-momentum 
tensor. This implies an effective gravitational field equation.
Scalar field parameters determine a cosmological constant (aka dark 
energy) in the resulting cosmological Friedmann equation. 
Empirical values of the relevant parameters can be fitted to redshift 
data implying Hubble expansion and acceleration\cite{MAN06}.
Thus in conformal theory the scalar field,
postulated to generate gauge boson mass in the standard model, is also 
responsible for cosmological expansion.  Qualitatively, this
is not surprising, since the Higgs mechanism renormalizes the universal
vacuum state, producing a nonvanishing scalar field amplitude $\phi_0$ 
throughout the universe.  Its energy-momentum tensor is a 
cosmological entity.    
\par Following sign conventions of electroweak theory\cite{CAG98,PAS95},
diagonal metric tensor $g_{\mu\nu}$ has elements $(1,-1,-1,-1)$ in flat 
space.  A covariant energy-momentum tensor
\begin{eqnarray}
\Theta_a^{\mu\nu}=\frac{-2\delta I_a}{\sqrt{-g}\delta g_{\mu\nu}}.
\end{eqnarray}
is determined by any invariant action integral
$I_a=\int d^4x \sqrt{-g} {\cal L}_a$.
This defines $\Theta_a^{00}$ as an energy density.
Conformal symmetry implies that $\Theta_a^{\mu\nu}$ is traceless.  For
gravitational Lagrangian density ${\cal L}_g$, functional derivative
\begin{eqnarray}
X_g^{\mu\nu}=\frac{\delta I_g}{\sqrt{-g}\delta g_{\mu\nu}}
\end{eqnarray}
implies the gravitational field equation
\begin{eqnarray}
X_g^{\mu\nu}=\half\sum_a \Theta_a^{\mu\nu}.
\end{eqnarray}
\par If $\delta{\cal L}_g=x_g^{\mu\nu}\delta g_{\mu\nu}$,
up to a 4-divergence, the functional derivative of action integral
$I_g$ is $X^{\mu\nu}_g=x_g^{\mu\nu}+\half{\cal L}_g g^{\mu\nu}$.
Standard Einstein-Hilbert theory, with Ricci scalar 
$R=g_{\mu\nu}R^{\mu\nu}$, cosmological constant $\Lambda$,
and ${\cal L}_g=(R-2\Lambda)/2\kappa$, 
implies the Einstein field equation
\begin{eqnarray}
G^{\mu\nu}+\Lambda g^{\mu\nu}=-\kappa\sum_a \Theta_a^{\mu\nu},
\end{eqnarray}
where $G^{\mu\nu}=R^{\mu\nu}-\half Rg^{\mu\nu}$ is the Einstein tensor.
Its trace $G=-R$ does not in general vanish.
\par Uniform, isotropic cosmology is characterized by Robertson-Walker
(R-W) geometry.  With the present sign conventions, the metric tensor
is defined for spatial curvature $k$ by
\begin{eqnarray}
ds^2=dt^2
-a(t)^2(\frac{dr^2}{1-kr^2}+r^2d\theta^2+r^2\sin^2\theta d\phi^2).
\end{eqnarray}
The Ricci tensor $R^{\mu\nu}$ (contracted Riemann) 
depends on universal scale factor $a(t)$ through only
two independent functions, $\xi_0(t)=\frac{\ddot a}{a}$ and
$\xi_1(t)=\frac{{\dot a}^2}{a^2}+\frac{k}{a^2}$.
The Ricci scalar is $R(t)=6(\xi_0(t)+\xi_1(t))$.
In R-W geometry, functional derivative $X_g^{\mu\nu}$ vanishes
for the conformal gravitational Lagrangian\cite{MAN06}.
If averaged uniform matter and radiation produce
$\Theta_m^{\mu\nu}$, the field equation reduces to
$\Theta_\Phi^{\mu\nu}+\Theta_m^{\mu\nu}=0$.
\par Electroweak theory\cite{REN90,CAG98} postulates an SU(2) doublet
complex scalar field $\Phi$ whose Lagrangian density contains
\begin{eqnarray}
\Delta{\cal L}_\Phi=w^2\Phi^\dag\Phi-\lambda(\Phi^\dag\Phi)^2,
\end{eqnarray}
omitting a constant term.  If $\phi_0$ is a spacetime constant, and
$\phi_0^2=\frac{w^2}{2\lambda}$ for $\lambda>0$, $\Phi=\phi_0$ is an
exact global solution of the scalar field equation.  This determines a  
stable vacuum state.  $\Delta{\cal L}_\Phi$ has residual value
$\half w^2\phi_0^2$.
\par ${\cal L}_\Phi=
 (\partial_\mu\Phi)^\dag\partial^\mu\Phi-\sxth  R\Phi^\dag\Phi
 -\lambda(\Phi^\dag\Phi)^2$
defines a conformally invariant action integral\cite{MAN06}. Electroweak
theory adds a term $w^2\Phi^\dag\Phi$, which breaks conformal symmetry. 
In R-W geometry, the occurrence of Ricci scalar $R$ here implies a  
Friedmann cosmic evolution equation with coefficients determined by
the scalar field Lagrangian\cite{MAN06}.
$R$ varies on a cosmological time scale.  The resulting
time variation of $\phi_0$ is many orders of magnitude smaller than that
relevant to elementary particle physics.  The present analysis will
treat $R$ as a constant in the scalar field equation,
$\partial_\mu\partial^\mu\Phi= 
 (-\sxth R+w^2-2\lambda\Phi^\dag\Phi)\Phi$.
\par If constant $\phi_0$, such that $\phi_0^2=w^2/2\lambda$, is
substituted for the scalar field, as in the Higgs construction,
${\cal L}_\Phi=-\sxth\phi_0^2(R-3w^2)$.
This acts as an effective gravitational Lagrangian density,  
of the same form as Einstein-Hilbert, but with parameters
${\bar\kappa}=-\frac{3}{\phi_0^2},{\bar\Lambda}=\frac{3}{2}w^2$.
The implied field equation is 
$G^{\mu\nu}+{\bar\Lambda}g^{\mu\nu}
 =-{\bar\kappa}\Theta_m^{\mu\nu}$.
The gravitational constant is negative and the cosmological constant is
determined by scalar field parameters\cite{MAN06}.
\par Generalizing the Higgs construction, for $\phi_0$ such 
that $\phi_0^2=\frac{1}{2\lambda}(w^2-\sxth R)$, $\Phi=\phi_0$ is a
global solution of the scalar field equation. The time derivative of 
$R$ can be neglected in the scalar field equation.
The algebraic sign of $\lambda$ must agree with $w^2-\sxth R$.
Because ${\cal L}_\Phi$ depends on $R$, its functional derivative for
variation of the metric tensor is $X_g^{\mu\nu}= 
\sxth R^{\mu\nu}\Phi^\dag\Phi+\half{\cal L}_\Phi g^{\mu\nu}$.
Evaluated for $\phi_0^2=\frac{1}{2\lambda}(w^2-\sxth R)$, 
${\cal L}_\Phi=\phi_0^2(w^2-\sxth R-\lambda\phi_0^2)
=\sxth\phi_0^2(3w^2-\half R)
=\frac{1}{4\lambda}(w^2-\sxth R)^2$.
Hence the effective gravitational functional derivative is
$X_g^{\mu\nu}
=\sxth\phi_0^2(R^{\mu\nu}-\frth Rg^{\mu\nu}+\frac{3}{2}w^2g^{\mu\nu})$.
The middle term here is reduced by a factor of two from the
corresponding Einstein-Hilbert expression. Defining
${\bar\kappa}=-\frac{3}{\phi_0^2}$ and ${\bar\Lambda}=\frac{3}{2}w^2$,
the modified gravitational field equation is
$R^{\mu\nu}-\frth Rg^{\mu\nu}+{\bar\Lambda}g^{\mu\nu}
 =-{\bar\kappa}\Theta_m^{\mu\nu}$.

\section{The cosmological Friedmann equation}
In the R-W metric, $R^{00}=3\xi_0$ and $R=6(\xi_0+\xi_1)$. For
energy density $\Theta_m^{00}=\rho$, the effective Einstein field
equation implies standard Friedmann cosmic evolution equation 
$-\thrd G^{00}=\xi_1(t)=\frac{{\dot a}^2}{a^2}+\frac{k}{a^2}
 =\thrd({\bar\kappa}\rho+{\bar\Lambda})$,
expressed in terms of scalar field parameters.
The effective gravitational field equation derived above,
taking into account Lagrangian term $-\sxth R\Phi^\dag\Phi$ and
the modified Higgs construction, implies
$R^{00}-\frth R=3\xi_0-\frac{3}{2}(\xi_0+\xi_1)
    =-{\bar\kappa}\rho-{\bar\Lambda}$.
This reduces to a modified Friedmann equation
\begin{eqnarray}\label{Fried1}
\xi_1(t)-\xi_0(t)=\frac{{\dot a}^2}{a^2}+\frac{k}{a^2}-\frac{\ddot a}{a}
    =\frac{2}{3}({\bar\kappa}\rho+{\bar\Lambda}).
\end{eqnarray}
For positive energy density, ${\bar\kappa}\rho$ here is negative, 
compensated by ${\bar\Lambda}$, spatial curvature $k$,
and acceleration ${\ddot a}$. 
\par In the early universe, with no stable masses, conformal symmetry
requires trace $\Theta_m=g_{\mu\nu}\Theta_m^{\mu\nu}$  to
vanish. Trace $\Theta_\Phi$ must also vanish. Using
$g_{\mu\nu}(R^{\mu\nu}-\frth Rg^{\mu\nu})=0$, gravitational trace
terms cancel identically in the effective field equation, implying 
${\bar\Lambda}=0$. Then for $k\to0$ Eq.(\ref{Fried1}) reduces to
$\frac{d}{dt}\frac{{\dot a}}{a}=-\frac{2}{3}{\bar\kappa}\rho$, which
correctly implies exponential expansion because ${\bar\kappa}<0$ in
conformal theory.  This is consistent with the postulate of universal
conformal symmetry\cite{MAN06}, prior to dynamical symmetry-breaking,
and with the hypothesis that parameter $w^2$ is due to
self-interaction associated with such symmetry-breaking\cite{NES08}.
For temperatures below the electroweak transition temperature $T_{EW}$,
conformal symmetry is broken dynamically by the Higgs mechanism. 
In this epoch, trace $\Theta_m$ cannot be assumed to vanish. 
Once conformal symmetry is broken and stable nonzero mass
is possible, R-W geometry is modified by mass concentrations and
may not remain strictly valid as a cosmological model.

\section{Empirical parameters}
In the present conformal theory scalar field parameters $w^2,\lambda$
determine scalar amplitude $\phi_0$, cosmological constant  
${\bar{\Lambda}}$, and effective gravitational constant ${\bar\kappa}$.
In a Robertson-Walker metric, scale factor $a(t)$ defines expansion
rate $H(t)={\dot a}(t)/a(t)$, whose present value $H_0=H(t_0)$
is the Hubble constant.  The parametrized gravitational equations  
determine a Friedmann cosmic evolution equation which relates
${\bar{\Lambda}}$ and energy density $\rho$
to cosmic expansion $a(t)$ and acceleration
$\frac{{\ddot a}a}{{\dot a}^2}=-q(t)$.
Defining dimensionless quantities, and normalizing to $a(t_0)=1$ at
present time $t_0$,
\\$
\Omega_m=\frac{2{\bar\kappa}\rho}{3H_0^2};
\Omega_\Lambda=\frac{2{\bar\Lambda}}{3H_0^2}=\frac{w^2}{H_0^2};
\Omega_k=-\frac{k}{H_0^2};
\Omega_q=-q(t_0),
$\\
the modified Friedmann Eq.(\ref{Fried1}) reduces at $t_0$ to
\begin{eqnarray}
\Omega_m+\Omega_\Lambda+\Omega_k+\Omega_q=1.
\end {eqnarray}
\par Coefficient
${\bar\kappa }$ is unrelated to the Newton constant $8\pi G_N$, which
retains its validity for gravitational dynamics in the solar system.
The full conformal theory is required for nonuniform or nonisotropic
energy-momentum density on a galactic scale\cite{MAN06}.  The sign
reversal of ${\bar\kappa}$ relative to $G_N$ implies $\Omega_m\leq 0$. 
Empirical values of $\Omega_\Lambda$ are positive, while $\Omega_k$ is
near zero.  If ${\bar\Lambda}>0$ and $k<0$, the evolution equation
implies that the current value of $\Omega_m$ is very small\cite{MAN06}. 
Given $0<\Omega_\Lambda+\Omega_k<1$, empirically well-defined,
the residual parameter from the modified Friedmann equation is 
$1>\Omega_m+\Omega_q>0$, which is compatible with $\Omega_m\leq 0$.
In contrast, the standard dimensionless Friedmann
equation, $\Omega_m+\Omega_\Lambda+\Omega_k=1$ implies
$1>\Omega_m>0$, conflicting with conformal theory.   
Conclusions regarding $\Omega_m$ based on the standard equation may
require reconsideration.
\par The algebraic sign of parameter $\lambda$ must agree with 
$w^2-\sxth R$.  At $t_0$ the independent field parameters are
$\xi_0(t_0)=\Omega_q H_0^2$ and $\xi_1(t_0)=(1-\Omega_k) H_0^2$,
so that $\sxth R=\xi_0+\xi_1=(1-\Omega_k+\Omega_q)H_0^2$.
Thus $(w^2-\sxth R)/H_0^2=\Omega_\Lambda+\Omega_k-\Omega_q-1
 =-2\Omega_q-\Omega_m$, negative for $\Omega_m<0$ if 
$\Omega_q+\Omega_m>0$.  A consistent model with $\Omega_q+\Omega_m>0$
requires $\lambda<0$.   Because the resulting value of  
${\cal L}_g=\frac{1}{4\lambda}(w^2-\sxth R)^2$ is negative,
$\lambda<0$ does not destabilize the physical vacuum ground state. 

\section{Numerical solution of the modified Friedmann equation}
The modified Friedmann equation 
$\frac{{\dot a}^2}{a^2}-\frac{{\ddot a}}{a}=
 \alpha-\frac{k}{a^2}-\frac{\beta}{a^3}$ 
can be solved numerically, given constant parameters 
$\alpha=\frac{2}{3}{\bar\Lambda}>0$, $k\simeq0$, and
$\beta=-\frac{2}{3}{\bar\kappa}\rho a^3>0$.  For luminosity distance
$d_L$, $H_0d_L$ is computed as a function of redshift $z$.
By adjusting the free parameter $\Omega_q$, Mannheim\cite{MAN03}
fitted his Eq.(11), a solution of the standard Friedmann equation for
$k=0$ and $\beta=0$, to empirical redshift data.  The fitted function 
is indistinguishable from a standard Hubble plot of the same data, 
with consensus parameters $\Omega_\Lambda=0.7$, $\Omega_m=0.3$. 
Similar results are given here for the modified Friedmann equation,
also with vanishing $k$ and $\beta$.  Here $\Omega_q$ is determined by
the solution, while free parameter $\alpha$ is adjusted to match 
Mannheim's $H_0d_L$ for redshift $z\simeq 1$.  The functions 
agree to graphical accuracy for $z\leq 1$.  Remarkably, the fitted 
parameter $\Omega_\Lambda=\alpha=0.732$ for $\Omega_k=0$ is consistent 
with current empirical values $\Omega_\Lambda=0.726\pm0.015, \Omega_k=
-0.005\pm 0.013$\cite{KOM09}.  Any significant discrepancy would 
invalidate the present theory.
\begin{table}[h]
\begin{tabular}{lcccc}
z& $\Omega_\Lambda$&$\Omega_q$&$H_0d_L(calc)$
 &$H_0d_L$(\cite{MAN03})\\ \hline
0.000& 0.732& 0.268& 0.000& 0.000\\
0.063& 0.672& 0.328& 0.066& 0.066\\
0.133& 0.619& 0.381& 0.145& 0.145\\
0.211& 0.571& 0.429& 0.240& 0.241\\
0.298& 0.530& 0.470& 0.355& 0.357\\
0.395& 0.492& 0.508& 0.494& 0.497\\
0.503& 0.459& 0.541& 0.663& 0.666\\
0.623& 0.428& 0.572& 0.868& 0.871\\
0.758& 0.401& 0.599& 1.118& 1.121\\
0.909& 0.376& 0.624& 1.424& 1.426\\
1.079& 0.353& 0.647& 1.799& 1.799\\
\end{tabular}
\end{table}

\section{Relationship to standard electroweak theory}
\par Because the standard model Lagrangian for the SU(2) complex scalar
field omits Ricci scalar $R$, the implied scalar field equation has a
time-independent global solution.  If both $w^2$ and $\lambda$ are
positive, this implies Higgs symmetry-breaking.   
Conformal theory replaces $w^2$ by $w^2-\sxth R$, where
$R$ is time-dependent on a cosmological scale, as implied by the
modified Friedmann equation.  This extremely weak time-dependence
establishes a correspondingly small but nonzero coupling to the
$Z^0$ weak boson field, through the $SU(2)$ covariant derivative
that acts on the scalar field\cite{CAG98,REN90}.  If virtual emission
and absorption of the $Z^0$ field produces a dressed scalar field,
parameters $w^2$ and $\lambda$ would represent the implied induced 
self-interaction\cite{NES08}.  Because the $Z^0$ transition amplitudes 
depend on cosmological time-derivatives, parameter values would
be extremely small. 
\par The current empirical value of the Hubble constant is
$H_0\simeq 70.5$km/s/Mpc\cite{KOM09}. 
Using $\Omega_\Lambda=0.726$, the present theory implies Higgs mass
$m_H=\sqrt{2\Omega_\Lambda}H_0=1.81\times10^{-33}eV$.
The standard Friedmann equation implies essentially the same result, 
differing by a factor of order unity.

\section{Conclusions}
\par It has been shown here that requiring fundamental gravitational 
and scalar field action integrals to have conformal scaling symmetry,
well-established for massless spinors coupled to gauge boson fields,
provides plausible explanations of several puzzling phenomena in 
elementary particle and gravitational physics.  The most striking are 
the long-term failure to observe a massive Higgs boson, and the need in
empirical cosmology for dark energy or an equivalent cosmological 
constant\cite{MAN06}.
\par Conformal theory justifies an effectively unified gravitational
and elementary-particle theory.  The postulated SU(2) scalar boson 
field is the common element that links these traditionally incompatible
theories.  There is no need to quantize gravitational
theory or to geometrize quantum field theory.  A classical metric 
field serves as the blackboard upon which quantum theory is written.
\par Conformal theory relates gravitational and electroweak parameters
through the cosmological constant ${\bar\Lambda}=\frac{3}{2}w^2$.
The Higgs construction of electroweak theory, which assumes
$w^2>0$, establishes a nonzero cosmological constant, equivalent to
uniformly distributed dark energy.  This parameter can be attributed to
self-interaction of the scalar field, due to virtual excitation of the
$Z^0$ gauge field\cite{NES08}.  This process is effective only in the 
current epoch of cosmic evolution, when mean temperature is below the
electroweak transition temperature $T_{EW}$. 
\par If this is the correct explanation of empirical dark energy,
it implies an extremely small Higgs boson mass, found here to be
$m_H\simeq 10^{-33}eV$.  Although this value might appear to be 
unreasonably small, no heavier Higgs boson has been detected.
Arguments that exclude small Higgs mass\cite{GHKD90} depend on
Higgs-fermion coupling.  As has recently been shown\cite{NES08}, a 
modified symmetry postulate in the standard model removes such coupling
terms, while justifying a plausible estimate of neutrino mass. 
\par The author is indebted to Prof. P. D. Mannheim for introducing
him to conformal symmetry and for helpful comments. 

\end{document}